\documentclass[pra,aps,twocolumn,showpacs,floatfix]{revtex4}

\usepackage{mathrsfs}
\usepackage{graphicx}           
\usepackage{dcolumn}            
\usepackage{bm}                 
\usepackage{hyperref}           
\usepackage{hypernat}           
\usepackage[latin1]{inputenc}   
\usepackage{pstricks}           
\usepackage{amsmath,amssymb}
\usepackage{version}

\DeclareGraphicsExtensions{.eps,.ps}

\newcommand{\beqa}{\begin{eqnarray}}
\newcommand{\eeqa}{\end{eqnarray}}

\renewcommand{\Re}{{\rm Re}}

\begin{document}

\title
{Streaking and Wigner time delays in photoemission from atoms and
surfaces}
\author{C.-H. Zhang and U. Thumm}
\affiliation{Department of Physics, Kansas State University,
Manhattan, Kansas 66506, USA}
\date{\today}

\begin{abstract}

Streaked photoemission metrology allows the observation of an
apparent relative time delay between the detection of photoelectrons
from different initial electronic states. This relative delay is
obtained by recording the photoelectron yield as a function of the
delay between an ionizing ultrashort extended ultraviolet (XUV)
pulse and a streaking infrared (IR) pulse. Theoretically,
photoemission delays can be defined based on i) the phase shift the
photoelectron wavefunction accumulates during the release and
propagation of the photoelectron (``Wigner delay") and,
alternatively, ii) the streaking trace in the calculated
photoemission spectrum (``streaking delay").   We investigate the
relation between Wigner and streaking delays in the photoemission
from atomic and solid-surface targets. For solid targets and
assuming a vanishing IR-skin depth, both Wigner and streaking delays
can be interpreted as an average propagation time needed by
photoelectrons to reach the surface, while the two delays differ for
non-vanishing skin depths. For atomic targets, the difference
between Wigner and streaking delays depends on the range of the
ionic potential.

\end{abstract}


\maketitle

\section{Introduction}

Streaked photoemission spectroscopy is increasingly applied to
resolve ultra-fast electronic processes at the natural timescale
($\approx 1$ atomic unit $= 2.4\times10^{-17}$s $=24$ attoseconds
(as)) of the motion of valence electrons in matter. Streaking
metrology uses ultra-short pulses of extreme ultraviolet (XUV)
radiation to emit electrons into the electric field of a delayed
infrared (IR) laser pulse. The XUV and IR pulses in this pump-probe
setup are phase coherent, and the yield of emitted photoelectrons is
recorded as a function of the delay $\Delta t$ between the two
pulses~\cite{Krausz09}. The resulting energy-resolved photoemission
spectra show stripes that oscillate in $\Delta t$ with the period of
the IR-laser electric field. These streaking traces occur in
distinct photoelectron kinetic energy intervals that are determined
by the spectral width of the XUV pulse, the IR-laser intensity, and
the density of states of the target. For photoemission out of
energetically resolved discrete atomic levels, streaking traces can
be related to a given initial state, while for photoemission from
electronic states in solids they are modulated by the density of
states within a given band of occupied states. The analysis of
streaked photoemission spectra proceeds by fitting the center of
energy (COE) of a given streaking trace as a function of $\Delta t$
to a sine function with an adjustable phase, thereby mapping energy
shifts induced by the IR laser onto a time delay between the
apparent release of the photoelectron and its arrival of the XUV
pulse~\cite{Zhang09}.

Streaked photoelectron spectra from localized states of atomic
targets~\cite{Schultze10} and both, delocalized conduction band and
localized core-level bands of solids~\cite{Cavalieri07} have
recently been recorded, leading to an ongoing debate about i) the
interpretation of the deduced photoemission time delays and ii) the
possibility of distinguishing delay contributions from the primary
XUV-photorelease process and subsequent photoelectron propagation in
the ionic potential of the target and the streaking IR-laser
electric field~\cite{Kheifets10, Zhang10,Ivanov11,Nagele11}.
Applying streaking metrology to neon atoms, Schultze {\it et
al.}~\cite{Schultze10} measured a {\em relative} photoemission
streaking delay of $\Delta\tau_S = 21 \pm 5$~as for the release of
electrons from 2p orbitals relative to emission from 2s orbitals.
The authors theoretically analyzed their measured relative delay in
terms of the relative Wigner delay $\Delta\tau_W (\epsilon)$ that is
given by the energy derivative of the spectral phase of the
calculated photoelectron wave function~\cite{Wigner55,Smith60} and
averaged $\Delta\tau_W (\epsilon)$ over the spectral profile of the
XUV pulse. The difference between their calculated averaged relative
Wigner delays $\overline{\Delta\tau}_W$ for emission from the 2s and
2p orbitals did not exceed
$\overline{\Delta\tau}(2p-2s)=\overline{\tau}_W (2p) -
\overline{\tau}_W (2s) = 6.4$ as, even for calculations that
included electronic correlation in neon at the
multiconfiguration-Hartree-Fock level. The authors linked the
mismatch between their measured relative streaking and calculated
Wigner delays to the extreme sensitivity of the photoelectron
wavefunction's spectral phase to electronic correlation effects in
multielectron atoms.

The same experiment~\cite{Schultze10} was subsequently analyzed by
Kheifets and Ivanov~\cite{Kheifets10} based on numerical solutions
of the time-dependent Schr\"odinger equation (TDSE) for a single
active electron moving in the Hartree-Fock potential of a Ne$^+$ ion
and, in a separate approach, by including electronic correlation
effects to some extent by numerically solving a set of coupled
equations in a random-phase approximation-with-exchange model. These
calculations reproduce only less than one half of the measured
relative delay of $21$~as, and the authors speculated that the much
larger observed relative delay might not be due solely to the
XUV-induced release process, even if electronic correlation effects
were accurately accounted for. The measured relative streaking delay
might thus include significant contributions from the
photoelectron's interaction with the streaking IR-laser electric
field. Indeed, the single-electron TSDE calculations by
Ivanov~\cite{Ivanov11} showed that the IR electric field has a
considerable influence on the Wigner delay for photoemission. The
question then arises whether this IR-dressed Wigner delay can be
used to interpret the measured relative streaking delay. As shown in
our previous investigation of time-resolved photoemission from a
one-dimensional (1D) model hydrogen atom~\cite{Zhang10}, the
streaking delay is independent of the streaking laser field
intensity. This was confirmed recently in a full-dimensionality
calculation for atoms by Nagele {\em et al.}~\cite{Nagele11}.

Investigating photoemission from a tungsten surface, Cavalieri {\it
et al.}~\cite{Cavalieri07} have measured a relative streaking delay
of $\Delta\tau_S(CB-4f) = \tau_S(CB) - \tau_S(4f) = 110 \pm 70$~as
for electrons emitted from 4f core levels relative to electrons
released from the conduction band. This relative delay was
interpreted as the delayed onset of IR streaking, i.e., as the
difference in time needed by 4f and conduction band electrons to
reach the surface~\cite{Cavalieri07,Kazansky09}. However, as we will
argue in Sec.~\ref{subsec:solid} below, this interpretation is only
valid under the assumption that the streaking IR field is fully
screened inside the solid. We will also show in this work that
Wigner and streaking delays become identical only in this limit of a
sudden onset of IR streaking at the surface. We will show that the
streaking delay sensitively depends on the IR-skin depth
$\delta_{L}$. Therefore, the intuitive interpretation of relative
streaking delays in terms of an effective photoelectron path length
inside the solid becomes questionable for realistic values of
$\delta_{L}$, depending on how exactly the IR electric field becomes
screened in the solid.

While only relative streaking delays can be deduced from measured
photoemission spectra, Wigner time delays are conveniently derived
from calculated photoelectron wave functions. Wigner and streaking
delays in time-resolved atomic photoemission were examined
recently~\cite{Ivanov11,Nagele11}, and the nature of delays within
the general context of scattering, decay, and photo- and
particle-induced emission processes in atomic, nuclear, other
branches of physics has been discussed by theorists for more than
half a century~\cite{Wigner55,Smith60} (for a recent review
see~\cite{Carvalho02}.) In this work, we  investigate the relation
between Wigner and streaking delays for photoemission from atoms in
the gas phase and solid surfaces. In Sec.~ \ref{sec:Wig_str}, we
present the underlying theoretical models and our schemes for
calculating time delays in photoemission. In Sec.~\ref{sec:results},
we compare and discuss our numerical results of time-resolved
photoemission spectra from atoms (Sec. \ref{sec:results}A) and
core-level and conduction bands of solid targets (Sec.
\ref{sec:results}B). In particular, we investigate the dependence of
the corresponding time delays on the XUV-photon energy, the
effective range $z_c$ of the atomic model potential, initial state,
IR-skin depth, and position of the Fermi level. Our conclusions
follow in Sec.~IV. Unless indicated otherwise, we use atomic units
(a.u.) throughout this work.

\section{Definition and computation of Wigner and streaking time delays}
\label{sec:Wig_str}

The essence of the time delay introduced by Wigner and
Smith~\cite{Wigner55,Smith60} can be understood for the elementary
example of potential scattering in one spatial dimension, by
representing the projectile as an incident wave packet
\begin{align}
\delta\psi_{in}(z,t)=\int dka_ke^{ikz-i\varepsilon_kt}
\end{align}
in terms of a superposition with amplitudes $a_k$ of plane waves
with momenta $k$ centered about $k_c$ and energies
$\varepsilon_k$. Scattering subject to the
projectile-target-interaction potential $V$ of finite range results
in the outgoing wave
\begin{align}
\delta\psi_{out}(z,t)=\int
dka_ke^{i\varphi_k}e^{ikz-i\varepsilon_kt}
\end{align}
for which each spectral component is phase shifted by $\varphi_k$
relative to the corresponding component of the incident wave.
Depending on the values of the scattering phase shifts $\varphi_k $,
wave fronts and center of the scattered wave are shifted relative to
the incident wave. The phase shifts  $\varphi_k$ thus quantify the
effect of $V$ on $\delta\psi_{in}$. Depending on the nature of $V$,
wave fronts of all plane-wave components and the center and crest of
the scattered wave packet may appear behind or ahead of the
corresponding terms of the incident wave packet. The scattered wave
packet can thus be characterized by a positive or negative delay
time $\tau$, depending on whether its wave fronts or center are
detected after or before they would be detected in the absence of
$V$, respectively. More precisely, the phase shifts of individual
traveling plane wave components lead to spectral delays
\begin{align}
\label{eq:wi-spectral} \tau_W(\varepsilon_k) =
\frac{\partial\varphi_k}{\partial \varepsilon_k},
\end{align}
which, evaluated at the spectral center $\varepsilon_c = k_c^2/2$ of
the incident wave packet, define the Wigner
delay~\cite{Wigner55,Smith60}
\begin{align}
\label{eq:wi} \tau^I_W=\tau_W(\varepsilon_c).
\end{align}

\begin{figure}[t]
\begin{center}
\includegraphics[width=1.0\columnwidth,keepaspectratio=true,
draft=false]{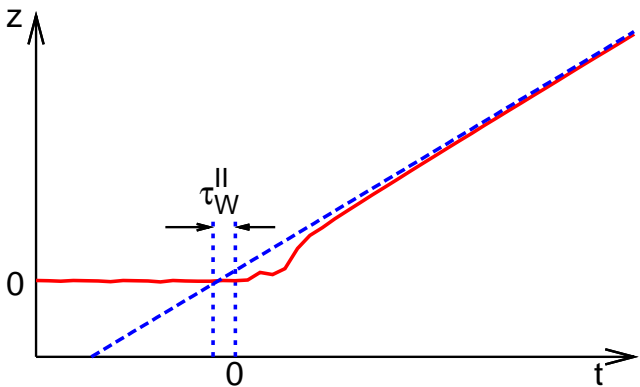}  \vspace{-6mm} \caption{(Color online)
Interpretation of the Wigner delay in the photoionization of atoms.
The solid line shows schematically the expectation values for the
position $\langle z\rangle$ of the photoelectron wave packet. The
photoemission Wigner delay $\tau_{W}^{II}$ is determined by a
straight-line extrapolation of $\langle z\rangle$ according to
Eq.~(\ref{eq:wii}) (dashed line). The XUV pulse is centered at
$t=0$. \label{fig:traject} } \vspace{-6mm}
\end{center}
\end{figure}

An alternative method for assessing the delay of the scattered
relative to the incident wave packet is given in terms of the
expectation values for the position,
\begin{align}
\langle z\rangle(t)=\int_0^{\infty}dzz|\delta\psi_{out}(z,t)|^2,
\end{align}
and velocity,
\begin{align}
\langle v\rangle=\int_0^{\infty} dkk\delta\phi_{out}(k,t)|^2,
\end{align}
of the scattered wave packet at a sufficiently large time $t>T$
after the interaction according to~\cite{Brenig59,Carvalho02}
\begin{align}
\label{eq:wii} \langle z\rangle=\langle v\rangle
(t-\tau^{II}_W),
\end{align}
where $\delta\phi_{out}(k,t)$ is the Fourier transformation of
$\delta\psi_{out}(z,t)$. The time $T$ is is chosen so that
$V(\langle z\rangle(t>T))\sim 0$ and all spectral phases $\varphi_k$
and $\langle v\rangle$ of the outgoing wave packet remain
time-independent to a very good approximation for $t>T$. This
fitting procedure can be understood classically by identifying the
expectation values for position and velocity in Eq.~(\ref{eq:wii})
with the motion of a point particle. We refer to both, $\tau^{I}_W$
and $\tau^{II}_W$, as ``Wigner delay" since it can be
shown~\cite{Carvalho02} that they are closely related  by the
expression
\begin{align}
\label{eq:wiia} \tau^{II}_W&=\int dk|\delta\phi_{out}(k,t)|^2
\tau^I_{W}(\varepsilon_k).
\end{align}
Indeed, our numerical results in Sec. \ref{sec:results} below
confirm that $\tau^{I}_W$ and $\tau^{II}_W$ are almost identically.

Both definitions of the Wigner time delay, $\tau^{I}_W$ and
$\tau^{II}_W$, have been used to characterize the scattering
processes~\cite{Carvalho02} and can also be applied to determine
delays in photoemission. They were recently used to interpret the
relative delay in the IR streaked XUV spectra from the 2s and 2p
shells in neon~\cite{Schultze10,Kheifets10,Ivanov11}. We thus
believe that it is important to carefully investigate the relation
between the Wigner and streaking delays.

We calculate the two Wigner delays for photoionization by solving
the TDSE for photoelectron wave packets $\delta\psi(z,t)$ emitted
from an initial state $\psi_i(z,t)$~\cite{Zhang10},
\begin{align}
\label{eq:ex} i\frac{\partial}{\partial
t}\delta\psi(z,t)=&\left[-\frac{p^2}{2}+V(z)\right]\delta\psi(z,t)
\displaybreak[0]\nonumber\\
&+zE_X(t)\psi_i(z,t),
\end{align}
where $p=i d/dz$ is the momentum operator and
$\psi_i(z,t)=e^{i|\varepsilon_B|t}\psi_i(z)$ the stationary initial
state with binding energy $\varepsilon_B$. For convenience we drop
the subscript ``out" and designate the outgoing photoelectron wave
simply as $\delta\psi(z,t)$. We consider photoemission by an
attosecond XUV pulse from either the ground state of a 1D model atom
or from the energetically lowest bands of occupied initial states in
the periodic 1D model potential of a solid. For the atomic case, we
choose $\psi_i(z,t)$ to be the ground state of the model atom with a
potential $V(z)$. For photoemission from a surface, $\psi_i(z,t)$
designates Bloch waves from within a given band with Block momenta
$k_i$.

We represent the coupling of the XUV pulse electric field $E_X$ to
the electron in the dipole-length form and assume a Gaussian pulse
profile,
\begin{align}
E_X(t)\sim e^{-2\ln2(t/\tau_X)^2}\sin(\omega_X t).
\end{align}
We assume a pulse duration $\tau_X=300$~as and a variable central
frequency $\omega_X$.

We propagate the photoelectron wave packet from $-4$~fs to $8$~fs on
a spatial grid that extends over 32,000~a.u. and calculate the
Wigner delays $\tau^{I,II}_W$ according to (\ref{eq:wi}) and
(\ref{eq:wii}). Rewriting the photoelectron wave packet at the large
time $T \gg \tau_X$ as~\cite{Zhang10,Ivanov11}
\begin{align}
\label{eq:pe_wave} \delta\psi(z,t=T)&\sim \int\!\!
dk\psi_{k}(z)d_{k}\tilde{E}_X(\varepsilon_k-\varepsilon_B)
e^{-i\varepsilon_kT},
\end{align}
where $\psi_k(z)$ is a continuum eigenstate in the potential $V(z)$
with energy $\varepsilon_k$, $d_{k} = \langle\psi_k
|z|\psi_i\rangle$ the dipole matrix element, and
$\tilde{E}_X(\omega)=\int dt E_X(t)e^{i\omega t}$ the spectrum of
$E_X(t)$, demonstrates that the Wigner delays depend on both, the
electron release during the dipole coupling of the XUV electric
field and the propagation of the photoelectron in the continuum.

In order to calculate streaking delays from the streaked
photoemission spectra, the influence of the streaking IR laser field
on the active electron need to be investigated. The IR streaking
effect on the release and propagation of the photoelectron is
included by replacing $p$ with $p+A_L(z,t-\Delta t)$ in
Eq.~(\ref{eq:ex}). The influence of the IR-laser on the initial
state can be included by numerically propagating the initial state
in the IR-laser electric field according to~\cite{Zhang10},
\begin{align} \label{eq:init}
i\frac{\partial}{\partial t}\psi_i(z,t)
=&\left\{\frac12\left[p+A_L(z,t-\Delta t)\right]^2
+V(z)\right\}\psi_i(z,t),
\end{align}
where $\Delta t$ is the delay between the centers of the XUV and IR
pulses, and the convention is used that $\Delta t>0$ corresponds to
the XUV pulse preceding the IR pulse. We model the vector potential
of the IR-laser pulse as
\begin{align}
A_{L}(t)=A_{0}\sin^2\left(\pi
t/\tau_L\right)\cos\left[\omega_{L}\left(t-\tau_L/2\right)\right]
\end{align}
for $0\le t\le \tau_L$ and set $A_L$ to 0 otherwise. As pulse
parameters, we choose the central photon energy
$\hbar\omega_{L}=1.57$~eV (corresponding to a wavelength of
$\lambda_L= 800$~nm), peak intensity $I_{L}= A_0^2\omega_L^2/2=
5\times10^{11}$W/cm$^2$, and pulse length $\tau_{L}=8$~fs.

Assuming a free-electron dispersion $(\varepsilon=k^2/2)$, the
energy-differential photoemission probability is given by
\begin{align}
\label{eq:probtau} P(\varepsilon,\Delta t)=\frac{1}{k} \left|
\delta\phi(k,\infty;\Delta t)\right|^2,
\end{align}
where $\delta\phi(k,\infty;\Delta t)$ is the Fourier transform of
$\delta\psi(z,t\rightarrow\infty;\Delta t)$. The XUV-IR
delay-dependent COE for a given streaking trace is~\cite{Zhang09}
\begin{align}
\label{eq:Ecom} E_{COE}(\Delta t)=\frac{1}{2P_{tot}(\Delta
t)}\!\!\int \!\!dk\left|k \, \delta\phi(k,\infty;\Delta t)\right|^2,
\end{align}
with the total emission probability
\begin{align}
\label{eq:Prob} P_{tot}(\Delta t)=\int dk
\left|\delta\phi(k,\infty;\Delta t)\right|^2.
\end{align}
After calculating $E_{COE}(\Delta t)$ for a range of XUV-IR delays
$-\tau_L/2\le\Delta t\le \tau_L/2$, we obtain the streaking delay
$\tau_S$ relative to $A_L$ by fitting the parameters $a$, $b$, and
$\tau_S$ to the expression~\cite{Zhang09,Zhang10}
\begin{align}
\label{eq:Ecom_fit} E_{COE}(\Delta t)=a+bA_L(\Delta t-\tau_S).
\end{align}

For XUV photoemission from solids,
Eqs.~(\ref{eq:probtau})-(\ref{eq:Ecom_fit}) remain valid. In this
case, the initial states in Eq.~(\ref{eq:ex}) are individual Bloch
waves with momenta in the first Brillouin zone of either core-level
or conduction band. We will show in Sec.~\ref{subsec:solid} below
how to calculate band-averaged results.

\section{Numerical results}
\label{sec:results}

\subsection{One-dimensional model hydrogen atom}
\label{subsec:atom}

\begin{figure}[t]
\begin{center}
\includegraphics[width=1.0\columnwidth,keepaspectratio=true,
draft=false]{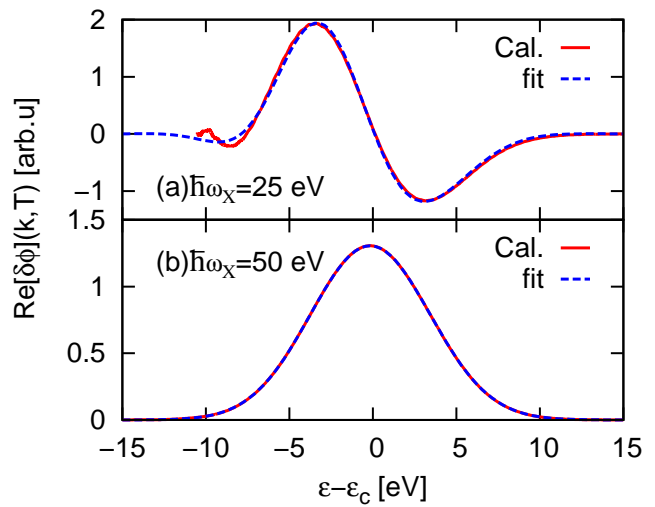}  \vspace{-6mm} \caption{(Color online) Real
part of the calculated and the fitted momentum-space photoelectron
wave packets for ionization from 1D model hydrogen atoms by XUV
pulses with energies of (a) 25~eV and (b) 50~eV. \label{fig:phase} }
\vspace{-6mm}
\end{center}
\end{figure}

In this section, we discuss our numerical results for Wigner and
streaking delays for XUV photoemission from the ground state with
binding energy $\varepsilon_B=13.6$~eV of the soft-core Coulomb
potential
\begin{align} \label{eq:SCPOT}
V(z)=V_c(z)=-1/\sqrt{z^2+2}.
\end{align}
In the calculation of the Wigner delay $\tau_W^I$, the direct
numerical determination of the phase of the photoelectron wave
packet according to $\varphi_k=\ln\delta\phi(k,T)/|\delta\phi(k,T)$
is inaccurate or impossible for  values of $k$ where
$|\delta\phi(k,T)|$ is extremely small. Furthermore, $\varepsilon_c$
is difficult to determine from $\delta\phi(k,T)$ [see
Fig.~\ref{fig:phase} (a)]. In order to overcome these two
difficulties, we fit the real part $\Re[\delta\phi(k,T)]$ of the
calculated photoelectron wave packet to the function
\begin{align}
\label{eq:Fit-f}
f(\varepsilon)=&Ae^{-2\ln2\left[(\varepsilon-\varepsilon_c)/\Delta
\varepsilon\right]^2}\cos(\varphi_k)
\end{align}
with
\begin{align}
\label{eq:Fit-phase}
\varphi_k=\alpha(\varepsilon-\varepsilon_c)+\beta(\varepsilon-\varepsilon_c)^2+\gamma,
\end{align}
and determine (by least-squares fit over a large range of
$\varepsilon=k^2/2$ values) the parameters $\varepsilon_c$,
$\Delta\varepsilon$, $\alpha=\tau^I_W$, $\beta$ and $\gamma$. As
shown in Fig~\ref{fig:phase} for photoionization with
$\hbar\omega_X=25$ and 50~eV,  the spectra of the calculated (by
numerically solving the TDSE) and fitted photoelectron wave packets
are in excellent agreement. Since according to
Eq.~(\ref{eq:pe_wave}) the XUV pulse spectral profile is imprinted
on the photoelectron wave packet, we find that the fitted values
$\Delta \varepsilon(\hbar\omega_X=25$~eV)=5.96~eV and $\Delta
\varepsilon(\hbar\omega_X=50$~eV)=6.11~eV are close to the spectral
width $\hbar\Delta\omega_X=6.08$~eV of the XUV pulse.

\begin{figure}[t]
\begin{center}
\includegraphics[width=1.0\columnwidth,keepaspectratio=true,
draft=false]{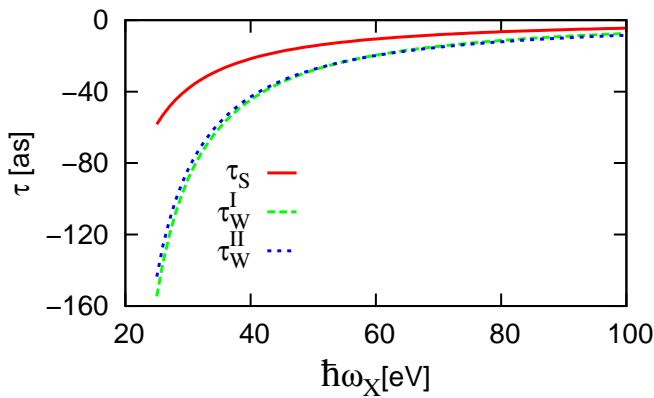}  \vspace{-6mm} \caption{(Color online)
Comparison of the Wigner delays $\tau^{I,II}_W$ and the streaking
delay $\tau_S$ for photoionization of 1D model hydrogen atoms as a
function of the XUV-photon energy $\hbar\omega_X$.
\label{fig:dtau_h1d_exuv} } \vspace{-6mm}
\end{center}
\end{figure}

\begin{figure}[t]
\begin{center}
\includegraphics[width=1.0\columnwidth,keepaspectratio=true,
draft=false]{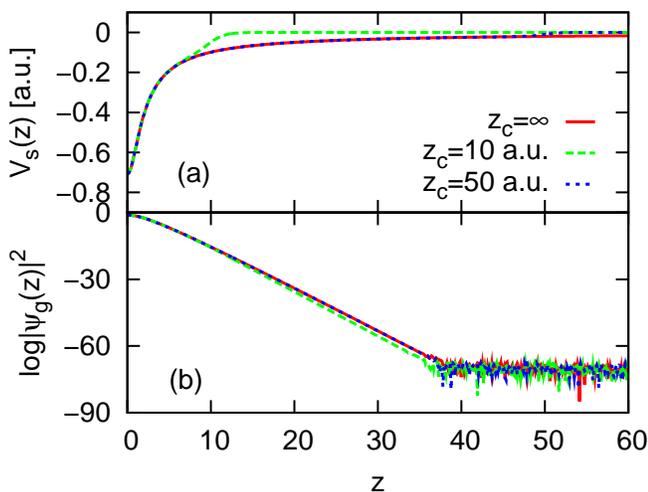}  \vspace{-6mm} \caption{(Color online) The
modified Coulomb potential (\ref{eq:vm}) (a) and its ground state
wavefunction (b) for three different interaction ranges $z_c$.
\label{fig:pot_short} } \vspace{-6mm}
\end{center}
\end{figure}

Figure~\ref{fig:dtau_h1d_exuv} shows the Wigner delays
$\tau^{I,II}_W$ for the 1D model hydrogen atom, calculated according
to Eqs.~(\ref{eq:wii}) and (\ref{eq:Fit-f})-(\ref{eq:Fit-phase}), in
comparison with the streaking delay $\tau_S$, calculated as
described in Sec.~\ref{sec:Wig_str}~\cite{Zhang10}. All delays are
negative. The two Wigner delays are almost identical and their
absolute values are much larger than the streaking delay. Having
established that the two Wigner delays almost coincide for all
parameters considered in this work, we only show results for
$\tau^{II}_W$ from now on and drop the superscript ``II" for
convenience, unless noted otherwise. In order to investigate the
difference between the Wigner and streaking delays, we modify the
infinite range of the potential Eq.~(\ref{eq:SCPOT}) with a
Wood-Saxon factor,
\begin{align} \label{eq:vm} V_s(z)=V_c(z)
\left[1-\frac{1}{1+e^{-(|z|-z_c)/a}}\right].
\end{align}
The range of $V_s(z)$ is controlled by  $z_c$. The parameter $a$
defines the lengths over which $V_c(z)$ is screened to approach
zero. In our calculation, we use $a=1$. For $z_c\rightarrow\infty$
$V_s$ converges to $V_c$ [Fig.~\ref{fig:pot_short}(a)]. We
numerically verified that for values of $z_c$ larger than a few
atomic units, the ground-state wavefunction and energy in $V_s(z)$
are practically independent of $z_c$ [Fig.~\ref{fig:pot_short}(b)].

\begin{figure}[t]
\begin{center}
\includegraphics[width=1.0\columnwidth,keepaspectratio=true,
draft=false]{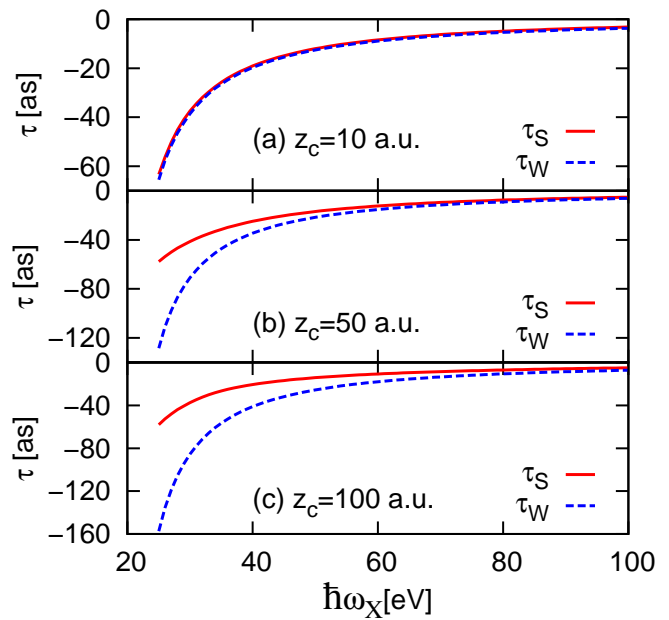}  \vspace{-6mm} \caption{(Color online)
Comparison of the Wigner delay $\tau_W$ and the streaking delay
$\tau_S$ for three interaction ranges $z_c$ in Eq.~(\ref{eq:vm}) as
a function of the XUV-photon energy $\hbar\omega_X$.
\label{fig:wigner_streak_short} } \vspace{-6mm}
\end{center}
\end{figure}

In Fig.~\ref{fig:wigner_streak_short} and Table~\ref{tab:delays} we
compare the Wigner and streaking delays at three different ranges
$z_c$. The comparison shows that for an interaction range slightly
larger than the extent of the ground-state probability distribution,
say, $z_d\sim2$, the Wigner and streaking delays coincide [see
Fig.~\ref{fig:wigner_streak_short} (a)]. In contrast, for $z_c\gg
z_d$, $|\tau_W|>|\tau_S|$, and both delays approach their values for
the 1D model potential Eq.~(\ref{eq:SCPOT}). In this case the
difference of the two delays is largest at lower photoelectron
kinetic energies [Fig.~\ref{fig:wigner_streak_short} (b) and (c)].
Comparison of the three graphs also shows that the Wigner delay is
more sensitive to changes in $z_c$ than the streaking delay.

\begin{table}[tbp]
  \begin{ruledtabular}
    \begin{tabular}{llll}
      $\hbar\omega_X$ [eV] & $z_c$ [a.u.]
        & $\tau_S $[as]
        & $\tau_W$ [as]\\
      \hline
      30 & 10 &  -37        &        -38\\
         & 50 &  -41        &        -71\\
         & 100&  -37        &        -85\\ \\
      50 & 10 &  -12        &        -12\\
         & 50 &  -17        &        -22\\
         & 100&  -14        &        -26\\ \\
      70 & 10 &  -6         &        -7 \\
         & 50 &  -9         &        -11\\
         & 100&  -8         &        -13\\
          \end{tabular}
  \end{ruledtabular}
\caption{Streaking and Wigner time delays from
Fig.~\ref{fig:wigner_streak_short} for three range parameters $z_c$
and three XUV-photon energies $\hbar\omega_X$. \label{tab:delays}}
\end{table}

We emphasize that Wigner delays are calculated without including the
action of an IR-laser pulse, which might give rise to the question
of whether we should instead compare the streaking delay with the
Wigner delay obtained in the same IR field, as mentioned in the
introduction. Obviously, inclusion of the IR-laser electric field
would change the Wigner delay~\cite{Ivanov11}. In this case the
photoelectron velocity in Eq.~(\ref{eq:wii}) would depend on the
XUV-IR delay $\Delta t$, and the use of this equation would
determine a Wigner delay $\tau_W^{II}$ that varies with $\Delta t$
and the IR-pulse intensity. The streaking delay, in contrast, does
not depend on $\Delta t$. It is also independent on the intensity of
the streaking laser, if this intensity is sufficiently
low~\cite{Zhang10,Zhang09,Nagele11}. Therefore, it is only
meaningful to compare streaking delays with IR-field-free Wigner
delays.

By computing photoemission spectra and streaking delays with and
without including the IR vector potential in Eq.~(\ref{eq:init}), we
found that for this atomic target polarization effects of the
initial state of the active electron in the electric field of the
streaking laser are negligible. This is due to the large energy gap
between the ground state and the excited states~\cite{Zhang10}. In
contrast, the initial state polarization is relevant for the case of
photoemission from solid surfaces discussed in the following
section.

\subsection{One-dimensional Solid}
\label{subsec:solid}

\begin{figure}[t]
\begin{center}
\includegraphics[width=1.0\columnwidth,keepaspectratio=true,
draft=false]{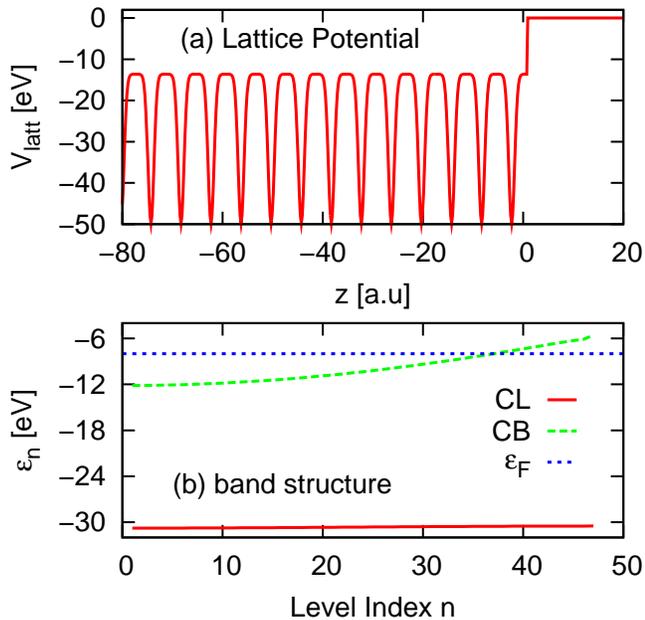}  \vspace{-6mm} \caption{(Color online) (a)
Model lattice potential consisting of equally-spaced Gaussian
potential wells. (b) Corresponding band structure with one
core-level band  and one conduction band. The dotted line indicates
Fermi level $\varepsilon_F$. \label{fig:pot_eigen} } \vspace{-6mm}
\end{center}
\end{figure}

We model the 1D solid surface as a row of $N$ equidistant atomic
layers and represent each  atom  by a Gaussian potential well to
form the lattice potential
\begin{align}
V_{latt}(z)=V_0-\frac{A_0}{\sqrt{2\pi}\sigma}\sum_{i=1}^{N}
e^{-[z+(i+0.5)a_{latt}]^2/(2\sigma^2)},
\end{align}
where $a_{latt}$ is the lattice constant, $\sigma$ controls the
overlap of the two adjacent atomic potentials, and $V_0$ and $A_0$
are chosen to match the known Fermi energy. We have oriented the
$z$-axis with increasing values towards the vacuum side and put the
origin ($z=0$) at the distance $0.5 a_{latt}$ in front of the top
nucleus. Diagonalizing the time-independent Schr\"odinger equation
\begin{align}
\label{eq:gs}
\varepsilon_n\psi_n(z)&=\left[-\frac{1}{2}\frac{d^2}{dz^2}+V_{latt}(z)\right]\psi_n(z)
\end{align}
for $N=47$, $V_0$=-0.5, $A_0$=2, $a_{latt}=6$, and
$\sigma=0.1a_{latt}$, we obtain the core-level and conduction-level
bands shown in Fig.~\ref{fig:pot_eigen}. The Fermi energy is
$\varepsilon_F=-10.9$~eV.

For the calculation of the XUV photoemission spectrum, we replace
$V(z)$ in Eq.~(\ref{eq:ex}) by $V_{latt}(z)$ and add the damping
term $-i v_z/(2\lambda)$, with the velocity
$v_z=\sqrt{2(\omega_X-|\varepsilon_n|)}$, to $V_{latt}(z)$, in order
to model  scattering of the photoelectrons inside the solid. In a
previous study, we adjusted the electron mean-free path inside the
solid to $\lambda=5$~\AA~as which corresponds to the minimum of the
universal curve for $\lambda$ as a function of the electron's
kinetic energy~\cite{Zangwill88}. We continue to use this value for
the present investigation. In the calculation of the streaked
spectra, we further assume an exponential damping of the IR-laser
field inside the solid,
\begin{align}
A_L(z,t)=A_L(t)\left[e^{z/\delta_L}\Theta(-z)+\Theta(z)\right],
\end{align}
characterized by the IR-skin depth $\delta_L$.

Within each band, we use the index $n$ to label Bloch wave functions
$\psi_n$ with energies $\varepsilon_n$, starting with
$\varepsilon_1$ for the lowest energy Bloch wave at the band bottom.
Each initial Bloch wave below the Fermi level contributes to the
photoemission spectrum with the energy-differential emission
probability
\begin{align} \label{eq:probepsilon-n}
P_n= P(\varepsilon_n,\Delta t)=\frac{1}{k} \left|
\delta\phi_n(k,\infty;\Delta t)\right|^2.
\end{align}
Similarly, each Bloch wave contributes to the COE $E_{COE,n}(\Delta
t)$, Wigner delays $\tau_{W,n}^{I,II}$, and streaking delays
$\tau_{S,n}$. We first calculate the band-averaged COE
\begin{align}
E_{COE}(\Delta t) = \frac{1}{\sum_n P_n}
\sum_{\varepsilon_n<\varepsilon_F} P_n E_{COE,n}(\Delta t)
\end{align}
separately for each band. Next we use Eq.~(\ref{eq:Ecom_fit}) to
obtain the band-averaged streaking delay $\tau_S$. Similarly, we
find the band-averaged Wigner delays according to
\begin{align}
\tau_{W}^{I,II}= \frac{1}{\sum_nP_n}
 \sum_{\varepsilon_n<\varepsilon_F} P_n \tau^{I,II}_{W,n}.
\end{align}

\begin{figure}[t]
\begin{center}
\includegraphics[width=1.0\columnwidth,keepaspectratio=true,
draft=false]{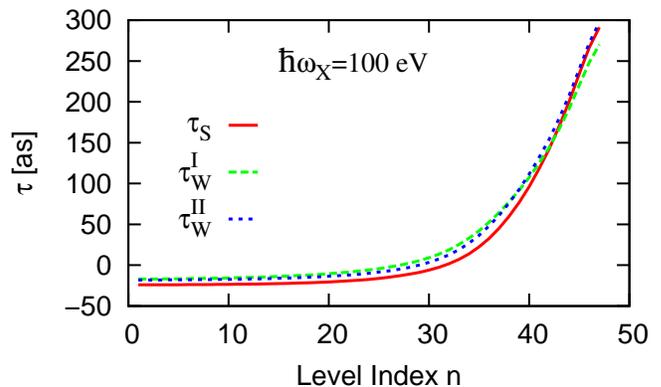}  \vspace{-6mm} \caption{(Color online)
Streaking and Wigner delays, $\tau_S$ and $\tau_{W}^{I,II}$, for XUV
photoemission with $\hbar\omega_X = 100$~eV from the core-level band
of a 1D model solid surface as a function of the core-level index
$n$. All delays are computed for the electron mean-free path
$\lambda=5$~\AA~ and no penetration of the IR-laser field into the
solid ($\delta_L=0$) for the streaking delay. \label{fig:dtau_n} }
\vspace{-6mm}
\end{center}
\end{figure}

A comparison of streaking and the two Wigner delays for an
XUV-photon energy $\hbar\omega_X = 100$~eV and emission from the
core-level band is shown in Fig.~\ref{fig:dtau_n} as a function of
the core-level index $n$. The monotonic increase in level index $n$
in Fig.~\ref{fig:dtau_n} is just a coincidence. For other energies,
these delays do not necessary increase with $n$. As for the atomic
case (cf., Sec.~\ref{sec:results}A), we find that the difference
between the two Wigner delays is negligible. We therefore only
present results for $\tau^{II}_{W}$ which we denote simply as
$\tau_W$ below. All numerical results shown below are converged in
the number of included atomic layers $N$.

\begin{figure}[t]
\begin{center}
\includegraphics[width=1.0\columnwidth,keepaspectratio=true,
draft=false]{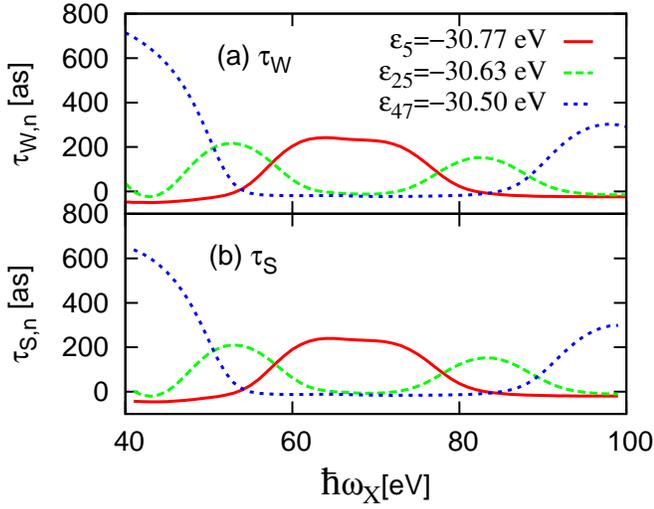}  \vspace{-6mm} \caption{(Color online) Wigner
(a) and streaking (b) time delays for XUV photoemission from three
core-level Bloch states with energies $\varepsilon_{5},
\varepsilon_{25}$, and $\varepsilon_{47}$ given relative to the
ionization limit. Streaking delays are computed for the electron
mean-free path $\lambda=5$~\AA~ and no penetration of the IR-laser
field into the solid ($\delta_L=0$).
\label{fig:dtau_solid_l5d0exuvclik} } \vspace{-6mm}
\end{center}
\end{figure}

\begin{figure}[t]
\begin{center}
\includegraphics[width=1.0\columnwidth,keepaspectratio=true,
draft=false]{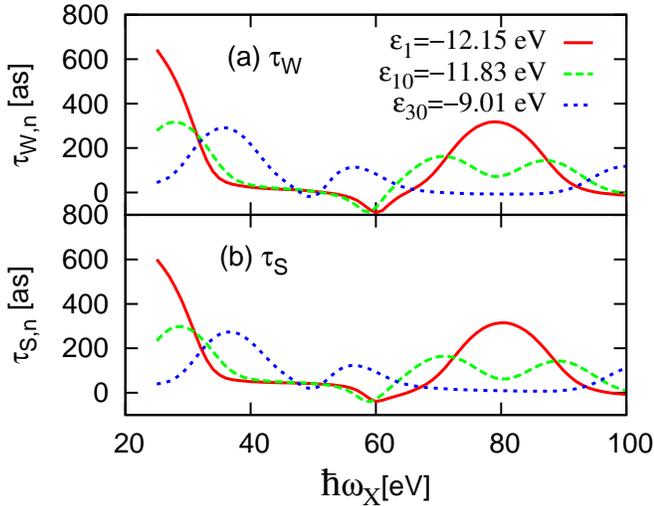}  \vspace{-6mm} \caption{(Color online) Wigner
(a) and streaking (b) time delays for XUV photoemission from three
conduction-band levels with energies $\varepsilon_1,
\varepsilon_{10}$, and $\varepsilon_{30}$ given relative to the
ionization limit. The occupied part of the conduction band extends
from $\varepsilon_1$ to the Fermi level at $\varepsilon_F=-8.22$~eV.
Streaking delays are computed for the electron mean-free path
$\lambda=5$~\AA~ and no penetration of the IR-laser field into the
solid ($\delta_L=0$).\label{fig:dtau_solid_l5d0exuvcbik} }
\vspace{-6mm}
\end{center}
\end{figure}

\begin{figure}[t]
\begin{center}
\includegraphics[width=1.0\columnwidth,keepaspectratio=true,
draft=false]{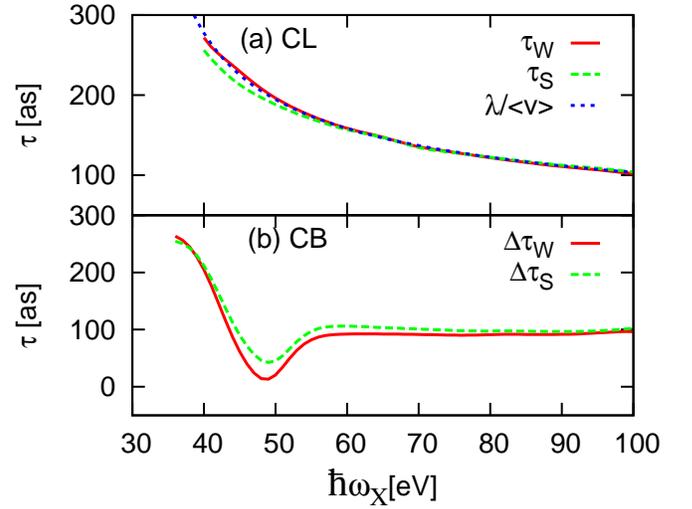}  \vspace{-6mm} \caption{(Color online) (a)
Band-averaged Wigner and streaking delays for XUV photoemission from
the (a) conduction band and (b) core-level band. Streaking delays
are computed for the electron mean-free path $\lambda=5$~\AA~ and no
penetration of the IR-laser field into the solid ($\delta_L=0$).
\label{fig:dtau_solid_l5d0exuv} } \vspace{-6mm}
\end{center}
\end{figure}

Wigner and streaking delays as a function of the XUV-photon energy
$\hbar\omega_X$ for photoemission from three individual core- and
three conduction-band Bloch levels are shown in
Figs.~\ref{fig:dtau_solid_l5d0exuvclik} and
\ref{fig:dtau_solid_l5d0exuvcbik}, respectively. According to its
definition Eq.~(\ref{eq:wii}), the Wigner delay can be regarded as
an ``effective" propagation time for the photoelectron to emerge
from the solid. For special case $\delta_L=0$, the streaking delay
is the travel time photoelectrons need before getting exposed to the
streaking IR field outside the solid. Therefore, intuitively, for
$\delta_L=0$ only, one would expect the streaking delay to be almost
identical to the Wigner delay. As
Figs.~\ref{fig:dtau_n}-\ref{fig:dtau_solid_l5d0exuv} show, this is
confirmed by our numerical results.

It is interesting to observe that, for emission from the core-level
band, the band-averaged Wigner and streaking delays decrease
monotonously with increasing $\hbar\omega_X$
[Fig.~\ref{fig:dtau_solid_l5d0exuv} (a)]. This decrease closely
follows the effective propagation time $\lambda/\sqrt{2\varepsilon}$
of photoelectrons inside the solid prior to reaching the
solid-vacuum interface at $z=0$, even though delay contributions
from individual Bloch levels, $\tau_{n}$ (see
Fig.~\ref{fig:dtau_solid_l5d0exuvclik}), do not show this behavior.
This confirms the interpretation that band-averaged Wigner and
streaking delays for emission from the core-level band can be
regarded as an average time needed for a released photoelectron to
travel a distance $\lambda$ inside the
solid~\cite{Cavalieri07,Kazansky09}.

However, this interpretation is not valid for photoemission from the
conduction band, where the band-averaged delays behave
non-monotonously as a function of $\hbar\omega_X$ as shown in
Fig.~\ref{fig:dtau_solid_l5d0exuv}(b). A possible explanation for
this difference is the delocalized nature of the conduction-band
Bloch wave. This can be checked by examining the core-level-
band-averaged delay as a function of the overlap parameter $\sigma$
in $V_{latt}(z)$ and will be discussed in a forthcoming
publication~\cite{overlap}. This means that, even for $\delta_L=0$,
the relative delays between photoemission from the core-level and
conduction band can not be due solely to the photoelectron's average
travel time in the solid~\cite{Cavalieri07,Kazansky09}.

\begin{figure}[t]
\begin{center}
\includegraphics[width=1.0\columnwidth,keepaspectratio=true,
draft=false]{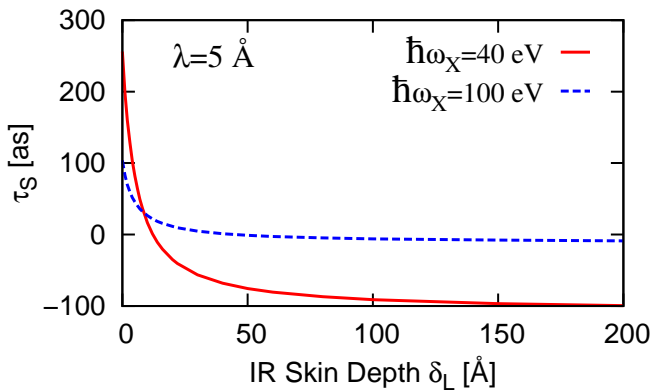}  \vspace{-6mm} \caption{(Color online)
Band-averaged streaking delays for XUV photoemission with photon
energies of $\hbar \omega_X=50$ and $100$~eV as a function of the
IR-laser skin depth  $\delta_L$. The electron mean-free path is
$\lambda=5$~\AA. \label{fig:dtau_delta_exuv} } \vspace{-6mm}
\end{center}
\end{figure}

Since the i) equivalence of $\tau_S$ and $\tau_W$ and ii) the
interpretation of streaking delays in terms of an effective
propagation time in the solid are only valid for the special case
$\delta_L=0$, we next investigate the dependence of photoemission
delays on the IR-skin depth $\delta_L$.
Figure~\ref{fig:dtau_delta_exuv} shows the band-averaged streaking
delay for emission from the core-level band for two XUV-photon
energies. These results show a very sensitive dependence of $\tau_S$
on the IR penetration depth, with $\tau_S$ changing from positive to
negative delays. In contrast to photoemission from the energetically
isolated ground states of atoms (Sec.~\ref{subsec:atom}), the $N$
Bloch waves form a quasi-continuum and can be easily hybridized in
the IR-laser electric field. This initial-state hybridization effect
is the stronger the deeper the IR electric field penetrates the
solid and accounts for the $\delta_L$ dependence of $\tau_S$. We
note that the actual IR-skin depth is much larger than the electron
mean-free path, such that the effective depth over which
photoelectrons are assembled is limited by $\lambda$, and the
IR-skin depth tends to become irrelevant for the
photocurrent~\cite{Zhang09}. Accordingly, the photoemission delays
in Fig.~\ref{fig:dtau_delta_exuv} converge in the limit of large
$\delta_L / \lambda$.

\begin{figure}[t]
\begin{center}
\includegraphics[width=1.0\columnwidth,keepaspectratio=true,
draft=false]{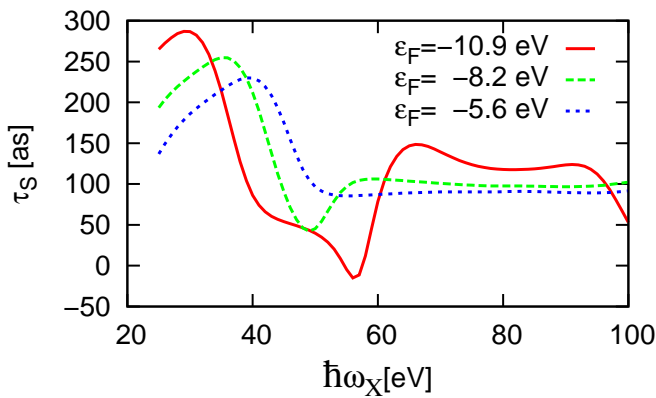}  \vspace{-6mm} \caption{(Color online)
Band-averaged streaking delay in the photoelectron spectrum for
photoemission from the conduction band for three values of the Fermi
energies $\varepsilon_F$. $\lambda=5$~\AA~ and $\delta_L=0$ are
used. \label{fig:Ecom_exuv25} } \vspace{-6mm}
\end{center}
\end{figure}

Another interesting observation, as shown in
Fig.~\ref{fig:Ecom_exuv25}, is that the band-averaged $\tau_W$ and
$\tau_S$ from the conduction band depend on the position of the
Fermi level. Therefore, changing the occupation probability of Bloch
waves in the conduction band, for example, by increasing the
temperature or by doping, will change the band-averaged delay. This
dependence is absent in the core-level band because all the core
levels are below the Fermi level and fully occupied.

\section{Conclusions}

We have examined the relation between Wigner and streaking delays in
the XUV photoelectron emission from model atoms and solid surfaces.
We showed that both, the creation of the photoelectron and its
propagation contribute to the Wigner delay. For photoemission from
atoms, the two delays are only identical for short-range ionic
potentials. For photoemission from  surfaces, Wigner and streaking
delays become identical only in the limit of no IR-field penetration
into the solid, and both delays can be interpreted as the travel
time of the photoelectron to the surface. Furthermore, for electron
emission from the core-level band, both delays can be understood as
the average time photoelectrons need to travel a distance equal to
the mean-free path in the solid. This interpretation does not hold
for photoelectron emission from the conduction band. This
dissimilarity is expected to be due to the different (localized
versus delocalized) nature of core and conduction-band levels.

We find that streaking delays are very sensitive to changes in the
IR-skin depth and Fermi energy and deviate from Wigner delays for
non-zero IR-skin depths. Their dependence on the substrate
temperature, impurities, and adsorbate coverage may leave a
measurable signature in relative photoemission delays.

\begin{acknowledgments}
This work was supported by the NSF, the Division of Chemical
Sciences, Office of Basic Energy Sciences, Office of Energy
Research, US~DOE. The computing for this project was performed on
the Beocat cluster at Kansas State University.
\end{acknowledgments}

\bibliographystyle{apsrev}
\bibliography{attosecond_full}

\end{document}